\documentclass[12pt,preprint]{aastex}

\input{psfig.sty}

\def\mo     {{$M_{\odot}$}}

\usepackage{epsfig,ref,natbib}
\bibpunct[]{(}{)}{,}{a}{}{,}

\begin{document}

\shorttitle{Mid-IR Statistics}
\shortauthors{Uzpen et al.}

\title{The Frequency of Warm Debris Disks 
and Transition Disks in a Complete Sample of Intermediate-Mass GLIMPSE Stars:
Placing Constraints on Disk Lifetimes}

\author{B. Uzpen \altaffilmark{1}, H. A. Kobulnicky \altaffilmark{1}, \and
K. Kinemuchi \altaffilmark{2} \altaffilmark{3}}


\altaffiltext{1}{University of Wyoming, Dept. of Physics \&
Astronomy, Dept. 3905, Laramie, WY 82071}

\altaffiltext{2}{Departamento de Astronom\'{i}a, Universidad de Concepci\'{o}n, Casilla 160-C,
Concepci\'{o}n, Chile}

\altaffiltext{3}{ Department of Astronomy, University of Florida,
211 Byrant Space Science Center, Gainesville, Florida 32611-2055, USA}

\begin{abstract}
The incidence of dusty debris disks around low- and intermediate-mass
stars has been investigated numerous times in order to understand the
early stages of planet formation.  Most notably, the $IRAS$ mission
observed the entire sky at mid- and far-IR wavelengths, identifying
the first debris disk systems, but was unable to detect a
statistically significant sample of warm debris disks due to its
limited sensitivity at 12 $\mu$m.  Using Tycho-2 Spectral Catalog
stars previously shown to exhibit 8 $\mu$m mid-infrared circumstellar
excesses confirmed at 24 $\mu$m in the \textit{Spitzer} GLIMPSE
survey, we investigate the frequency of mid-IR excesses among
intermediate-mass (2--4 \mo) stars in a complete volume-limited
sample. Our study of 338 stars is four times larger than a complete
sample of 12 $\mu$m sources from the $IRAS$ Point Source Catalog. We
find that 0.3$\pm$0.3$\%$ of intermediate-mass stars exhibit a
signature of a possible terrestrial-temperature debris disks at
wavelengths of 8 $\mu$m and greater. We also find that 1.2$\pm$0.6$\%$
of intermediate-mass stars exhibit evidence for circumstellar disks
undergoing inner disk clearing, i.e., candidate transition disk
systems.  Using stellar lifetimes and the frequency of transition and
primordial disks within a given spectral type, we find that
pre-main-sequence disks around intermediate-mass stars dissipate in
5$\pm$2 Myr, consistent with other studies.
\end{abstract}


\section{Introduction}

While it is generally accepted that most stars are surrounded by
primordial gaseous circumstellar accretion disks when they are formed
(e.g., \citealt{Waters:1998}), and that they gradually disperse, some
evolving toward dust-only ``debris disks" containing processed grains
(\citealt{Backman:1993}), there are gaps in understanding how these
two types of disks are connected.  From color-color studies of the
``Big Four'' debris disks (Vega, Fomalhaut, $\epsilon$ Eridani, and
$\beta$ Pictoris) it has been noted that debris disks may form a
continuum from very dusty systems, such as $\beta$ Pictoris, to less
dusty systems such as Vega (\citealt{Backman:1987};
\citealt{Decin:2003}). By studying the evolution of circumstellar
disks and determining the relative frequency of each evolutionary
stage, one may place constraints on the timescales of planet
formation. In this paper we conduct such a study of late-stage disks
around a large sample of intermediate-mass (2--4 \mo) stars---a
population lacking, until recently, sufficient representation in
infrared surveys to draw statistical conclusions.

Studies of individual star forming regions and stellar clusters have
concluded that primordial disks around intermediate-mass stars
dissipate more rapidly than their low-mass counterparts (e.g.,
\citealt{Dahm:2007}). The timescales for this dissipation vary from
less than 3 Myr (\citealt{Hernandez:2005}) up to $\sim$5 Myr
(\citealt{Carpenter:2006}), while low-mass objects may take as long as
10 Myr \citep{Sicilia-Aguilar:2006}.  However, most cluster studies
contain only a few to a few tens of intermediate-mass stars, rendering
the estimates for disk clearing timescales in this mass range
uncertain.  The time it takes for circumstellar disks to dissipate is
critically related to planet formation. The two most prominent planet
formation theories propose drastically different time scales for the
formation of planets, with core-accretion (\citealt{Goldreich:1973};
\citealt{Mizuno:1980}) requiring up to 8 Myr
(\citealt{Mordasini:2007}) and gravitational disk instability
(\citealt{Kuiper:1951}) requiring less than 0.1 Myr \citep{Boss:1997}.
Constraining the time necessary for disk clearing can test the
validity of planet formation theories.  Disk dissipation timescales of
much less than 3 Myr would imply rapid planetary formation and lend
support to the disk instability theory or require a refinement of the
core-accretion theory.

Toward the end of the primordial disk phase, the gaseous circumstellar
disk begins to dissipate, and with it, the signature of the infrared
excess. Primordial disks are characterized by large circumstellar
excesses at wavelengths greater than 1 $\mu$m indicated by their
fractional infrared luminosities (L$_{IR}$/L$_{*}$) greater than
10$^{-2}$ (\citealt{Hillenbrand:1992}). Later-stage disks, however,
have smaller fractional infrared luminosities and do not exhibit
near-IR excesses ($<$ 5 $\mu$m), which indicates that the inner
portion of the disk is largely devoid of material. Such late-stage
disks with mid-IR (5--12 $\mu$m) excesses and inner holes, but
containing primordial material at more distant radii, have been termed
''transition disks" (\citealt{Strom:1989}; \citealt{Strom:2005}).
Transition disks were once thought to be rare, and their rarity
implied rapid circumstellar disk dissipation (e.g.,
\citealt{Wolk:1996}). A majority of known transition disks surround
low-mass ($\sim$1 \mo\ FGK-type) stars \citep{Najita:2007}, for
example, TW Hya \citep{Calvet:2002}.  Recently, many new low-mass
transition disks have been identified, and they may comprise
$\sim$10\% of the pre-main-sequence population \citep{Cieza:2007}.  By
comparison, only a few candidate transition disks have been identified
around intermediate-mass stars (\citealt{Hernandez:2006};
\citealt{Hernandez:2007}; \citealt{Uzpen:2008}). The lack of a large
sample of transition disks around intermediate-mass stars may imply
different circumstellar evolutionary paths for different mass stars,
or may reflect the difficulty in observing transition disks in a
statistically meaningful sample of intermediate-mass stars.

Debris disks are the final chapter in the circumstellar evolutionary
sequence.  As a second generation disk of processed dust generated by
grain growth and planetesimal collisions, debris disks have fractional
infrared luminosities less than 10$^{-2}$, with most exhibiting
fractional infrared luminosities less than 10$^{-4}$
\citep{Artymowicz:1996}.  Debris disks were first identified by the
$\textit{Infrared Astronomical Satellite}$ ($IRAS$) on the basis of an
excess emission at $\lambda$ $>$ 25 $\mu$m detected during routine
calibrations of Vega \citep{Aumann:1984}. Subsequently, many debris
disks have been identified through the comparison of optical catalogs
with the $IRAS$ catalog (e.g., \citealt{Rhee:2007}).  At least 15\% of
nearby A--K main-sequence stars have dusty debris disks that were
detectable with \textit{IRAS} and $\textit{Infrared Space
Observatory}$ (\textit{ISO}) sensitivities in the far-infrared
(\citealt{Meyer:2006}; \citealt{Lagrange:2000};
\citealt{Backman:1993}). \citet{Plets:1999} found the disk fraction
(the fraction of systems exhibiting disks) for main-sequence stars and
their descendants to be 13$\pm$10\% and 14$\pm$5\%, respectively, at
60 $\mu$m. Investigating 160 A-type stars, \citet{Su:2006} found that
32$\pm$5$\%$ of their sample exhibits an excess at 24 $\mu$m. More
recently, \citet{Hillenbrand:2008} determined that $\sim$10$\%$ of
0.6--1.8 \mo\ main-sequence stars exhibit 70 $\mu$m emission
characteristic of debris disks. Debris disks may be indicators of
planetary systems \citep{Zuckerman:2004}, and \citet{Beichman:2005}
found that 24$\pm$10$\%$ of the stars exhibiting emission at 70 $\mu$m
due to dust also have planetary systems, while
\citet{Moro-Martin:2007} found that $\sim$10$\%$ of FGK stars have
planets regardless of disk presence. Debris disks identified to have
planets are both cool (T $<$ 200 K) and 10--100 times more massive
than the debris within our solar system \citep{Beichman:2005}. No
planet has yet been identified in a debris disk system exhibiting
shorter wavelength infrared excesses ($<$ 12 $\mu$m) due, in part, to
the scarcity of such sources.

Studies using $IRAS$ and the $ISO$ have shown a systematic drop in
infrared excess with stellar age for debris disks
(\citealt{Spangler:2001}; \citealt{Habing:2001}). Observing 266 A-type
stars with the $\textit{Spitzer Space Telescope}$ ($Spitzer$),
\citet{Rieke:2005} found that excess emission decreases with age as
$t$$_{0}$/$t$ with $t$$_{0}$=150 Myr. \citet{Su:2006} found a similar
trend between fractional infrared luminosity and stellar age along
with evidence that the inner portions of debris disks clear more
rapidly than the outer portions. \citet{Bryden:2006} investigated 69
FGK main-sequence stars and found that 2$\pm$2$\%$ of stars exhibited
$L$$_{IR}$/{$L$$_{*}$ $>$ 10$^{-4}$ with one star of the sample
showing evidence of warm dust (T $>$ 200 K).  \citet{Moor:2006}
investigated 60 debris disks within 120 pc and found that nearly all
the stars with ($L$$_{IR}$/{$L$$_{*}$) $>$ 10$^{-4}$ are young with
ages less than 100 Myr.  These results are consistent with previous
studies that conclude disk clearing occurs on short time-scales, at
young ages, and form from the inside out (e.g., \citealt{Habing:2001};
\citealt{Decin:2003}; \citealt{Mamajek:2004} and references therein).

Mid-IR excess debris disk sources, such as $\beta$ Pictoris, appear to
be extremely rare \citep{Chen:2005}.  Stars that exhibit a
mid-infrared excess have characteristic disk temperatures of 200--1000
K. Dust with these temperatures surrounding intermediate-mass stars is
located between $\sim$1--25 AU, and such regions are analogous to our
asteroid belt.  By comparison, long-wavelength excesses ($>$ 24
$\mu$m) have cooler (T$\sim$100 K) dust located farther from the
central star and are more comparable to the Kuiper Belt.
\citet{Aumann:1991} investigated 548 nearby main-sequence stars using
12 $\mu$m measurements from \textit{IRAS} and found that only 2 stars
($<$ 0.5\%) exhibited genuine excesses confirmed from ground based
observations. \citet{Aumann:1991} also found that out of the 35 nearby
A-type stars, only $\beta$ Pictoris exhibited a strong 10 $\mu$m
excess, resulting in a disk fraction of less than 3$\%$.  Prior to
$Spitzer$, only three intermediate-mass main-sequence or
near-main-sequence stars exhibited a strong mid-IR excess: $\beta$
Pictoris, $\zeta$ Leporis, and HR 4796A (\citealt{Aumann:1991};
\citealt{Metchev:2004}).  Eta Cha and HD 3003 exhibit a weak 12 $\mu$m
excess in the $IRAS$ Faint Source catalog \citep{Mannings:1998} and
Vega exhibits a weak near-IR excess \citep{Absil:2006}. Prior to
$Spitzer$, only three intermediate-mass stars were known to be
surrounded by disks predominately composed of warm dust with disk
temperatures greater than 200~K: $\eta$ Cha, $\zeta$ Leporis, and HD
3003 (\citealt{Mannings:1998}; \citealt{Chen:2001};
\citealt{Rhee:2007}).  With the vastly superior sensitivity of
$Spitzer$, a number of $IRAS$ 60 $\mu$m excess debris disks have been
shown to have excesses at shorter wavelengths (8.5--13 $\mu$m;
\citealt{Chen:2006}). $Spitzer$ has also been used to identify a
number of new candidate warm (T $>$ 200 K) debris disk systems around
early type stars that exhibit excesses at 8 and 24 $\mu$m
(\citealt{Uzpen:2005}; \citealt{Uzpen:2007}; \citealt{Hernandez:2006};
\citealt{Currie:2008b}).

In this paper, we present the statistical detection rates for a
complete\footnote{We use the word ``complete'' to mean volume-limited
sample constructed by applying a Malmquist bias correction to a
flux-limited sample.}  survey of intermediate-mass (2--4 \mo) mid-IR
excesses of indeterminate age identified in the \textit{Spitzer}
Galactic Legacy Infrared MidPlane Survey Extraordinaire (GLIMPSE;
\citealt{Benjamin:2003}). The goal of this paper is to compare the
disk fractions of warm debris disks and transition disks in a sample
of field stars to a compilation of young cluster members and to
determine the timescales for pre-main-sequence disk dissipation around
intermediate-mass stars.  Some of these field stars may be young stars
of yet unidentified clusters since warm debris disks are most common
around stars of age 10--15 Myr \citep{Currie:2008a}.  In \S 2, we
discuss how we obtained our sample of warm debris and transition disks
drawn from optical, spectral, and infrared catalogs following the
results of \citet{Uzpen:2007,Uzpen:2008}.  In \S 3 we determine our
statistically complete working sample. We also examine the constraints
our survey places on planetary formation and circumstellar disk
evolution. In \S 4, we compare the constraints our sample places on
the longevity and frequency of transition disk systems and the
frequency of warm debris disks to other studies. We conclude, in \S 5,
that debris disks and transition disk systems are rare and may help
constrain circumstellar disk evolution scenarios.

\section{The Warm Debris Disk/Transition Disk Sample}

Our stellar sample draws upon the Tycho-2 Spectral Catalog
\citep{Wright:2003}, 2MASS all-sky survey \citep{Cutri:2003}, and
GLIMPSE Catalog (v1.0).  The Tycho-2 Spectral Catalog contains more
than 351,000 positional matches of stars with known spectral types
from major spectral catalogs, such as the Henry Draper Catalog, with
the Tycho-2 Catalog \citep{Hog:2000} that contains both $B$ and $V$
photometric data and high quality astrometric data. The GLIMPSE
project is one of six original \textit{Spitzer} Legacy
Programs. GLIMPSE mapped the Galactic Plane at 3.6, 4.5, 5.8, and 8.0
$\mu$m with sensitivities of 0.6, 0.6, 2, and 10 mJy, respectively.
The GLIMPSE Catalog identified over 3$\times$10$^{7}$ sources within
the survey region. In the GLIMPSE data reduction pipeline, $JHK$
photometry was obtained from 2MASS images, and the 2MASS sources were
cross-correlated with GLIMPSE sources. Our resultant complete working
sample has both spectral information and photometric measurements in
the $B$, $V$, $J$, $H$, $K$, [3.6], [4.5], [5.8], and [8.0]
bandpasses.

In \citet{Uzpen:2007}, we identify 22 intermediate-mass stars in the
Tycho-2 Catalog and GLIMPSE survey of luminosity class V or IV and
spectral type B8 or later that exhibit 8 $\mu$m mid-infrared
extraphotospheric excesses.  These excess sources were identified from
among more than 4,000 stars investigated. The sample contained 1,024
stars with luminosity class V or IV of which 493 were of spectral type
B8--A5. Six of the 22 stars with infrared excesses did not have
luminosity classifications in the Tycho-2 Spectral Catalog. Instead,
the luminosity classifications were derived from observations in
\citet{Uzpen:2007} and classified therein.  These six stars were
removed from the statistical analysis of our complete working sample,
in \S 3 and 4, because they lacked a luminosity classification within
the Tycho-2 Spectral Catalog. We confirmed the excess at 8 $\mu$m by
finding an excess at 24 $\mu$m for 11 of the 16 remaining stars in
\citet{Uzpen:2007,Uzpen:2008}. Ten of the 11 remaining stars were
observed as part of the sample in \citet{Uzpen:2008} to determine the
origin of their infrared excess. Four of the 10 stars exhibited
H$\alpha$ emission and were found to have an excess owing to free-free
emission. Four of the stars exhibited H$\alpha$ emission at a level
insufficient to explain their infrared excess owing to free-free
emission.  The excesses for these four sources were consistent with
blackbody emission from warm dust, and these sources were classified
as candidate transition disks. One star exhibited H$\alpha$ emission
and optical spectral features characteristic of Herbig AeBe (HAeBe)
stars and is a candidate class II protostar. One of the 10 sources,
G311.0099+00.4156, did not exhibit H$\alpha$ emission and is a
candidate warm debris disk. Table~\ref{des} lists the six stars
exhibiting excesses ascribed to warm dust. This table also shows
stellar parameters (effective temperature, rotational velocity,
H$\alpha$ equivalent width) and circumstellar parameters (disk
temperature and fractional infrared luminosity) from
\citet{Uzpen:2008}.  The $K$-[8.0] and [8.0]-[24.0] colors are taken
from \citet{Uzpen:2007}.  Other terms have been used to define disks
undergoing disk clearing, e.g., ''anemic disks" \citep{Lada:2006}.  We
use the term ''transition disk" to mean a dissipating disk with
evidence of circumstellar gas, a fractional infrared luminosity
between 10$^{-2}$--10$^{-4}$, and an infrared excess due to dust.  As
reported in \citet{Uzpen:2008}, transition disk systems do not exhibit
Paschen emission lines, may exhibit \ion{O}{1} $\lambda$8446 emission,
or \ion{O}{1} $\lambda$7772 in emission or absorption, do not exhibit
\ion{Fe}{2} emission, and must have an excess at 8 and 24 $\mu$m not
entirely explainable by free-free emission. Presumably, transition
disks contain some primordial gaseous material as evidenced by their
gaseous emission lines. For the purposes of this paper, transition
disks are part of the group of disks, including class II protostellar
disk systems, that comprise primordial (pre-main-sequence) disk
systems. Therefore, both transition and class II protostellar disk
systems will be used to determine timescales for primordial disk
clearing. Warm debris disks are stellar systems lacking gaseous
emission with dust temperature of 200--1000 K that exhibit infrared
circumstellar excesses at 5--12 $\mu$m and longer wavelengths.  To
summarize, our sample of 493 B8--A5 luminosity class IV-V stars from
GLIMPSE, 2MASS, and Tycho-2 Spectral Catalogs contains four transition
disks, one class II protostellar disk object, and one warm debris
disk.

\section{Approach to Constructing a Volume-Limited Sample}

Like any flux-limited survey, use of the GLIMPSE catalog is subject to
Malmquist bias \citep{Malmquist:1922}.  Generically, this effect
arises from an inherent dispersion of absolute magnitude,
$\sigma_{M}$, among a sample of similar stars such that more luminous
objects scatter into a flux-limited sample because they are detected
at greater distances (\citealt{Binney:1998}; \citealt{Gilmore:1990}).
For our study, this potentially results in stars with 8 $\mu$m
excesses preferentially being detected to larger distances than
non-excess stars, thereby biasing the disk fractions toward higher
values.  In general terms, the {\it measured mean absolute magnitude},
$\bar{M}_{mag}$, that would be observed in a flux-limited sample is
brighter than the {\it true mean absolute magnitude}, $\bar{M}_{vol}$,
by a small factor that depends upon $\sigma_M$ and $A(m)$, which is
the star count function $dN(m)/dm$.  Mathematically,
\citet{Gilmore:1990} show that

\begin{equation}
\bar{M}_{mag}=\bar{M}_{vol}-ln 10\times \sigma_M^{2}\frac{d log A(m)}{dm}.
\label{eq1}
\end{equation}

\noindent Similarly, in the absence of interstellar extinction, the
{\it measured mean apparent magnitude}, $\bar{m}_{mag}$, that would be
observed in a flux-limited sample is brighter than the {\it true
apparent magnitude}, $\bar{m}_{vol}$, in a volume-limited sample,

\begin{equation}
\bar{m}_{mag}=\bar{m}_{vol}-ln 10\times \sigma_M^{2}\frac{d log A(m)}{dm}.
\label{eq2}
\end{equation}

\noindent Defining a volume-limited sample for a group of
similar spectral type stars requires knowledge of $\bar{m}_{vol}$
(effectively the apparent magnitude limit of the survey) which we will
call $m_{lim}$ and $\bar{M}_{vol}$ (usually based upon stellar models
or calibrated standard spectral types).  The limiting radial distance
is then determined from the usual distance modulus,

\begin{equation}
r=10^{0.2\times(\bar{m}_{mag}-\bar{M}_{vol})+1}.
\label{eq3}
\end{equation}
 
Interstellar extinction at the wavelength of interest, $A_{\lambda}$,
constitutes an additional factor that will act to drive the
completeness limit toward {\it fainter} apparent magnitudes.  A survey
is volume-limited (i.e., complete) at magnitudes brighter than
$m_{comp} \equiv \bar{m}_{mag}+A_\lambda$.  Selecting all stars with
$m < m_{comp}$ yields a volume-limited sample within radial distance
$r$.

\subsection{Construction of the Complete Working Sample}

Both $\bar{M}_{vol}$ and $\sigma_M$ can be measured from a sample of
stars with well-determined distances, and we use the $Hipparcos$
\citep{Perryman:1997} parallaxes in the All-sky Compiled Catalogue of
2.5 million stars \citep{Kharchenko:2001} to estimate these values.
As a test of our analysis methods, we computed $\bar{M}_{vol}$ and
$\sigma_M$ for a sample of 352 K0V stars and compared our results to
those of \citet{Butkevich:2005} and \citet{Oudmaijer:1999} who
analyzed similar stars using the same $Hipparcos$ dataset.  We
rejected the most uncertain measurements by requiring that the
parallaxes be positive and have $S/N$ $\equiv$ $\pi$/$\sigma$$_{\pi}$
$>$ $3$, leaving 347 stars.  We used the listed Tycho-2 spectral types
and observed $B-V$ colors to reddening correct each star assuming a
standard interstellar extinction law and R=3.1.  The top portion of
Figure~\ref{dist} shows the distribution of absolute magnitudes for
the 347 remaining K0V stars. The histogram has a narrow, roughly
Gaussian core with broad, asymmetric wings, most notably on the bright
side. This asymmetry may be understood as the result of unresolved
binaries and misclassified giants. In an effort to cleanse the sample
of these possible contaminants, we remove stars more than $3\sigma$
from the mean using a sigma clipping algorithm.  The resultant sample
contains 208 stars and is shown by the dash-dot-dot histogram.  The
dashed line in the top plot of Figure~\ref{dist} is a Gaussian
distribution with mean of $\bar{M}_{vol}$=5.62$\pm$0.14 mag and a
dispersion of $\sigma_M=0.29$ mag that characterizes the remaining 208
stars.  By comparison, \citet{Butkevich:2005} found
$\bar{M}_{vol}=5.81\pm0.01$ and $\sigma_M=0.33\pm0.02$ for a sample of
45 K0V stars fainter than 4.5 mag carefully chosen to remove binaries
and possible spectral misclassifications.  \citet{Oudmaijer:1999}
found $\bar{M}_{vol}=5.69$ and $\sigma_M=0.40$ for a sample of 85 K0V
stars with distances less than 50 pc and $V$ magnitudes greater than
4.5.  In summary, we determine estimates for $\bar{M}_{vol}$ and
$\sigma_M$ that are consistent with those in the literature,
suggesting that the same approach will yield suitable values for
applying a Malmquist bias corrections to our B8V -- A5V
sample. However, all three studies of K0V stars result in slightly
brighter measured mean absolute $V$ magnitudes than the 5.9 mag
tabulated in \citet{Cox:2000}. These studies used by \citet{Cox:2000},
(i.e., \citealt{Schmidt-Kaler:1982}; \citealt{Johnson:1966};
\citealt{DeJager:1987}), are prior to the launch of
$Hipparcos$. Distances calculated from $Hipparcos$ were improved,
which may explain the slight discrepancies.  Regardless, the small
differences between the $Hipparcos$ determined $\bar{M}_{vol}$ values
($<$ 0.2 mag) have a small effect ($<$ 10$\%$) on the definition of our
volume-limited sample and subsequent conclusions.
    
Our similar analysis of 1,095 A0V stars with $Hipparcos$ parallactic
measurements yields $\bar{M}_{vol}=0.92\pm0.20$ and $\sigma_M=0.65$.
The lower plot in Figure~\ref{dist} shows the distribution of the full
sample and a Gaussian distribution having the same mean and dispersion
as the 772 stars that remain after outlier rejection.  We find that
the dispersion of absolute magnitudes is approximately twice as large
for early-type stars compared to late-type stars. This implies that
there is either an intrinsically greater dispersion of luminosity in
early-type stars or there is a greater uncertainty in the
classification of early-type stars.  Additional reddening
uncertainties may affect the intermediate-mass sample to a greater
extent than the lower-mass sample, leading to a larger dispersion.
The median distance for the K0V sample is 50 pc, while the A0V sample
median distance is 220 pc. Small deviations of our assumed reddening
law results in dispersions four times greater for early-type stars
compared to late-type stars. Our adopted value of $\sigma_M=0.65$ may,
therefore, be regarded as a conservative upper limit that, when used
to correct for Malmquist bias, yields a relatively small
volume-limited sample of intermediate-mass stars.

The parameter $A(m)$ describes the differential star count per
apparent magnitude bin, which in turn depends upon the space density
distribution and the stellar luminosity function.  Under the
simplifying assumption of a homogeneously distributed population
having a Gaussian luminosity function, \citet{Binney:1998} show that
the factor ($d$ log A(m)/$dm$)=($d$ ln A(m)/ ln 10)/$dm$ in
Equation~{\ref{eq2} reduces to 0.6.  At the distances probed by our
survey ($<$1 kpc) and at $|b|$ $<$ 1$^\circ$, all of our
intermediate-mass stars will have $|z|$ $<$20 pc, thereby falling within
the thin disk of the Milky Way \citep{Haywood:1997} where the
assumption of homogeneity is a reasonable one.  Equation~\ref{eq2}
then reduces to

\begin{equation}
\bar{m}_{mag}=\bar{m}_{vol}- \sigma_M^{2} \times 0.6 .
\label{eq4}
\end{equation}

For A0V stars, the final term in Equation~\ref{eq4} becomes $0.65^2
\times 0.6 = 0.25$, meaning that our sample becomes volume-limited
at apparent magnitudes 0.25 mag brighter than the nominal flux limit
of the least sensitive dataset required for inclusion in the survey.
We require inclusion in the Tycho-2 Catalog (99\% complete to $V$=11
\citep{Hog:2000}), the 2MASS Catalog (99\% complete to $K$=14.3
\citep{Cutri:2003}), and the GLIMPSE Catalog (99.5\% complete to 10
mJy at IRAC band 4).  The GLIMPSE limit is equivalent to [8.0]=9.52
mag.  For spectral types near A0 that typify our sample, the
completeness limit of the GLIMPSE [8.0] data is ultimately the
limiting factor in constructing our ``complete working sample'', to be
defined below.  

In order to construct our complete working sample we passed Kurucz
ATLAS9 model atmosphere spectra through digitized photometric
bandpasses using temperatures and surface gravities corresponding to
the nearest spectral type to determine $V$-[8.0] colors
\citep{Kurucz:1993}.  Since all of the stars lie within the solar
circle, we adopt the solar metallicity models. For an A0V star we use
the stellar model of temperature 9,500~K with surface gravity of log
$g$=4.0.  The resulting colors were then used to determine our
limiting magnitude.  For example, an A0V star was found to have
$V$-[8.0]=0.06. In the absence of reddening we found that an A0V
star with M$_{V}$=0.92 determined from the parallaxes above will reach
the [8.0] limit at 540 pc with $V\equiv m_{lim}$=9.58. Similarly, an
A5V star will reach the [8.0] limit at 520 pc with $V\equiv
m_{lim}$=9.89 and a B8V star will reach the [8.0] limit at 730 pc with
$V\equiv m_{lim}$=9.28.

As an average value for interstellar extinction, we adopt $A_V=1.5$
mag~kpc$^{-1}$.  Our assumed value of extinction does not affect our
distance limit.  Only the computed $V$ band limits are affected by
extinction, with greater extinction pushing our $V$ band limits to
fainter values.  Since $A_V$ is $\lesssim$1.0 mag for distances up to
$\sim700$ pc in our survey, it has a small effect on $m_{comp}$, our
completeness limit, which is always brighter than the $V$-band limits
of the Tycho-2 catalog. If A$_{V}$ exceeds 2.5--3 magnitudes then our
correction for Malmquist bias would require an iterative process
between distance completeness limits and m$_{comp}$ with reddening
affecting our distance limit.  We solve for $m_{comp}$ using,

\begin{equation}
 m_{comp} = \bar{m}_{mag} + 0.015 [mag~pc^{-1}] \times r [pc] .
\label{eq5}
\end{equation}

\noindent As mentioned above, an A0V star reaches the 10 mJy limiting
[8.0] flux equivalent of $V\equiv m_{lim}=9.58$ at a distance of 540
pc in the absence of extinction and Malmquist bias corrections.
However, $9.58 - 0.65^{2} \times 0.6$ implies $\bar{m}_{mag}$=9.33,
corresponding to a volume-limited distance of 480 pc.  After applying
a correction for extinction, we find that A0V stars at a distance of
480 pc have $m_{comp}=$ 10.05. Table~\ref{comp} lists the distances,
D$_{sp}$, and $m_{comp}$, at which each spectral subtype, B8V through
A5V, is volume-limited.  These distances and magnitudes define our
``complete working sample''.

In addition to our flux limit of 10 mJy at [8.0], our sample has a
brightness limit at [8.0] of 700 mJy, which is the saturation limit
for IRAC and corresponds to a $V$$\simeq$5 star. There are only two
luminosity class V stars brighter than $V$=5 in the complete GLIMPSE
working sample.  Both of these stars are of spectral type A5, and have
[5.8] and [8.0] measurements in GLIMPSE although they have quality
flags, and they are saturated in the [3.6] and [4.5] bandpasses. Since
we could determine whether these stars exhibit an excess at [8.0], we
included them in our sample. For an unreddened B8V star, this
saturation limit occurs at $V$$\sim$4.7 mag and 87 pc and for an A5V
star, the saturation limit occurs at $V$$\sim$5.3 mag and 58 pc.

\subsection{Detection and Constraints on Disk Longevity from the Complete Working Sample}

Using the method outlined above, we find that there are 338 stars
between B8--A5 and luminosity class V and IV in the volume where our
sample is complete for each respective spectral type within the
Tycho-2 and the GLIMPSE surveys.  Table~\ref{types} shows the number
of stars of luminosity class V (column 2) or IV (column 3) within the
complete GLIMPSE working sample at each spectral type.  Stars listed
as IV in our sample are either luminosity class IV or IV/V stars and
do not include III/IV stars from the Tycho-2 Spectral Catalog.
Table~\ref{types} shows that our sample is mostly comprised of B9 and
A0 stars (175) with just a small fraction later than A2 (36).

Column 9 of Table~\ref{des} shows that our one warm debris disk
candidate (G311.0099+00.4156), our one class II protostellar disk
(G321.7868+00.4102), and all four transition disk candidates
(G305.4232--00.8229, G307.9784--00.7148, G311.6185+00.2469,
G314.3136--00.6977) fall within the magnitude limits of the complete
GLIMPSE working sample. The implied detection rates are then
0.3$\pm$0.3$\%$ (1/338$\pm$$\sqrt[]1/338$) for warm debris disks and
1.2$\pm$0.6$\%$ (4/338) for transition disk candidates, and
1.5$\pm$0.7$\%$ for all pre-main-sequence disks (class II protostellar
and transition disks) surrounding intermediate-mass stars. The
lifetime of a B8 star (4\ \mo) is 150 Myr, whereas a B9 star (3.5\
\mo) is 220 Myr and an A1 star (2.7\ \mo) is 430 Myr
\citep{Siess:2000}. The fraction of systems with transition disks
times the stellar lifetime yields an empirical statistical estimate
for the duration of the transition disk phase.  Transition disks are
the end points of pre-main-sequence circumstellar disks, and including
them with the number of class II protostellar disks times the stellar
lifetime provides insight into the time necessary for
pre-main-sequence disks to dissipate.

We found 45 B8 stars within the complete working sample. Two of these
45 stars exhibited transition disk features for a detection rate of
4.4$\pm$3.1$\%$. Given that the lifetime of a B8 star is 150 Myr, our
detection rate implies a lifetime of 6$\pm$4 Myr
(4.4$\pm$3.1$\%$$\times$150 Myr) for transition disks or 4.4$\%$ the
lifetime of a B8 star. We identified one transition disk, out of 93 B9
stars within our sample, and that led to a detection rate of
1.1$\pm$1.1$\%$.  The implied lifetime of a B9 tranistion disk is then
2$\pm$2 Myr.  There are 56 A1 stars within the sample, and we found
only one transition disk yielding a detection rate of 1.8$\pm$1.8$\%$.
Using the lifetime of 430 Myr for an A1 star, we find that transition
disks persist for 8$\pm$8 Myr. The results for B8, B9, and A1 stars
are consistent with one another but do not provide strong constraints
on the transition disk duration.

Considering the whole sample, we have a detection rate of four out of
338 sources for transition disks and are able to place a greater
constraint on the transition disk lifetime. This is done for a range
of spectral types rather than one spectral sub-class. Most of our
stars are of spectral types A0 and B9. In order to better estimate the
longevity of transition disks, we approximate the mean lifetime of our
sample. Table~\ref{tageest} shows the number of stars for a given
spectral type and the calculated lifetime for that star based on
\citet{Siess:2000} evolutionary models of the nearest stellar mass to
the measured spectral type. We determine that the weighted mean
lifetime of our sample is 360 Myr using the lifetime and the number of
stars at a given spectral type within our sample. Given our detection
rate, we estimate the lifetime of transition disk systems to be
4$\pm$2 Myr (i.e., 4/338 $\times$ 360 Myr).  In \citet{Uzpen:2008} we
estimate the age of G307.9784--00.7148, a transition disk system, to
be 1--2 Myr, consistent with the lifetime for these types of
objects. Unfortunately, these results do not provide strong enough
evidence to show that disks do indeed undergo rapid inner disk
clearing on timescales less than 1 Myr \citep{Wood:2002}, nor do these
results lend support to either of the two leading planet formation
theories.  These results do, however, demonstrate that
terrestrial-temperature material around intermediate-mass stars is
quite rare. Systems that exhibit terrestrial-temperature dust are
ideal candidates for follow-up observations to identify Earth-like
planets.

If we include transition disks in our sample of pre-main-sequence
disks, we can make an estimate of the lifetime of primordial
circumstellar disks around intermediate-mass stars. Looking at our
sample as a whole, we identified five pre-main-sequence disks (4
transition and 1 class II) out of 338 sources. Using a weighted mean
lifetime of 360 Myr, we determine that primordial disks persist for
5$\pm$2 Myr.  This result is consistent with cluster studies where
disk clearing occurs within $\sim$6 Myr \citep{Haisch:2001}.  We
summarize our results in Table~\ref{sum} for the disk fraction of warm
debris, transition, and class II protostellar disks within the sample
population and the primordial disk dissipation timescale for each
restriction considered.

\subsection{The $V$$\leq$9.25 Sample}

While the Tycho-2 catalog is 99\% complete to $V=11.0$, the Tycho-2
Spectral catalog from which our sample is selected, contains spectral
type and luminosity class information for only a subset of these
stars.  The brightest stars have essentially complete spectral
classification, while classifications for stars fainter than $V\sim9$
become increasingly incomplete, and this potentially introduces a bias
in the sample.  Figure~\ref{sphist} is a histogram of Tycho apparent
magnitude, $m_{V_T}$, for the Tycho-2 Catalog, which contains 2.5
million stars and the Tycho-2 Spectral Catalog containing 351,000
stars.  Although, the Tycho-2 Catalog is complete to $V=11.0$, the two
histograms start to diverge in the 9.25--9.5 histogram bin. Taking the
ratio of star counts per bin in the Tycho-2 Spectral Catalog to the
Tycho-2 Catalog, we find that this ratio is 88\% for the 9.25--9.50
magnitude bin. This means that for stars in Tycho-2 Spectral Catalog
at magnitudes fainter than 9.25, we do not expect all the stars to
have spectral types yet determined.  Our results may be biased into
the sense that we require a star to have a spectral
classification. For example, if the fraction of disk-bearing
pre-main-sequence stars increases toward fainter magnitudes, our
results may be biased.

We investigate whether requiring a spectral type affects our results
by defining a revised sample with spectral information restricted to
objects brighter than $V=9.25$, the ''$V$$\leq$9.25 sample".  This
restricted sample is listed in Table~\ref{types2}.  This revised
sample contains 163 of the original 338 stars.  The warm debris and
class II protostellar disks are still detected in this smaller cropped
sample.  Thus, the resulting disk fraction for warm debris disks is
0.6$\pm$0.6$\%$. We now detect one A1 transition disk out of 19 stars,
yielding a fraction of 5.3$\pm$5.3\% and an age upper limit of
23$\pm$23 Myr. The one B9 transition disk falls outside the 9.25
magnitude sample limit. The detection rate for B8 stars is two out of
19 stars yielding a disk fraction of 11$\pm$7\% and an age upper limit
of 16$\pm$11 Myr. For this entire sample, the fraction of transition
disks is three out of 163 sources, or 1.8$\pm$1.1\%, yielding an age
range of 6$\pm$4 Myr.  The disk fraction for pre-main-sequence objects
is four out of 163 sources, 2.5$\pm$1.2$\%$, resulting in an upper
limit for disk clearing of 9$\pm$4 Myr. The revised magnitude
constraint yields a larger lifetime for transition disk systems and
also a larger uncertainty compared to our complete working sample.
However, the results are still consistent with our values from the
complete working sample, suggesting that the incompleteness of the
Tycho-2 Spectral Catalog does not introduce large biases.  Working
under the presumption that the spectral types identified consist of a
representative sample of intermediate-mass stars at our magnitude
limit, we use our complete working sample.

\subsection{The Malmquist Biased Sample}

Prior to correcting for Malmquist bias, our sample consisted of 493
stars of spectral types B8--A5 and luminosity class IV or V and we
will refer to this sample as the ''biased sample."  Within this sample
there were four transition disk systems, one warm debris disk system,
and one class II protostellar disk system leading to detection rates
of 0.8$\pm$0.4$\%$ for transition disks, 0.2$\pm$0.2$\%$ for warm
debris disks, and a primordial disk detection rate of
1.0$\pm$0.5$\%$. All three of these biased detection rates are
consistent with the complete working sample results. Using our
weighted mean stellar lifetime of 360 Myr, the biased detection
frequency of transition disks implies that this phase persists for
3$\pm$1 Myr. This result also implies primordial disks clear within
4$\pm$2 Myr and is consistent with stellar cluster studies (e.g.,
\citealt{Lada:2006}) that find most primordial circumstellar disks
clear within 5 Myr.

\subsection{Additional Analysis and Consistency Check}

Taken at face value, our findings indicate that transition disks can
persist longer than the $\sim$1 Myr predicted from observations and
seem to be more common than the lack of detections in previous
observational studies would suggest (\citealt{Skrutskie:1990};
\cite{Wolk:1996}; \citealt{Wood:2002}; \citealt{Cieza:2007}).  Using
our selection criteria, transition disks are more common than their
progenitors, HAeBe stars, contrary to expectations. However, this may
simply be the result of small number statistics.  We considered
whether some selection effect would preferentially exclude HAeBe stars
from our sample. While there is no explicit criterion excluding these
objects, their copious circumstellar material produces systematically
higher visual extinctions.  For example, the well-studied HAeBe stars
from \citet{Hillenbrand:1992} have extinctions and distances that
imply $A_V\gg1.5$ mag kpc$^{-1}$, suggesting that in many such
objects, circumstellar extinction dominates over interstellar
extinction.  Such a circumstellar component would cause HAeBe stars to
drop out of the Tycho-2 Spectral Catalog from which our complete
sample is drawn and thereby explain their deficit in our statistics.

As an independent means of estimating the relative frequencies of
HAeBe stars, transition disk systems, and their main-sequence
descendants, we used 2MASS near-IR and GLIMPSE mid-IR colors to
identify objects of each evolutionary stage within the GLIMPSE survey
area.  The top plot of Figure~\ref{2color} shows a $J$--$H$
vs. $H$--$K$ diagram. This color space is commonly used to distinguish
pre-main-sequence stars (HAeBe stars) from their main-sequence
counterparts \citep{Hernandez:2005}. We take the unreddened color box
of \citet{Hernandez:2005} Figure~2 and apply A$_{V}$=10 to the upper
bounds of that box in order to define a region where we would expect
to find reddened HAeBe stars. Nearly all of the HAeBe stars from
\citet{Hillenbrand:1992} fall within this box.  Therefore, we assume
that this region will include most HAeBe stars.  Most lower-mass
pre-main-sequence stars (T-Tauri) from \citet{Hartmann:2005} fall
outside the HAeBe region and are not a major source of
contamination. Our transition disk candidates, by definition lacking
near-IR excesses, and $\beta$ Pictoris, a well-known debris disk
system, fall outside the HAeBe region and lie near the main-sequence.
We define a hexagonal region in this figure to estimate where most B
and A main-sequence stars can be found. The boundaries of the hexagon
are chosen to encompass unreddened B8--A5 main-sequence stars and
reddened B8 stars with A$_{V}\leq$4.  Extending the hexagon further to
the upper right (toward larger extinction) would begin to include
substantial numbers of reddened low-mass main-sequence stars and
possible T-Tauri stars. The bottom plot of Figure~\ref{2color} shows a
$Spitzer$ IRAC [4.5]--[8.0] vs. $J$--$K$ color-color diagram. This
color space is useful for distinguishing transition disk candidates
from other types of stars. The symbols are the same as those in the
upper plot of Figure~\ref{2color}.  The rectangular region defines
objects with [8.0] excesses where we find most transition disks. We
know from \citet{Uzpen:2008} that this region is contaminated by some
mid-IR excess debris disks and classical Be stars. However, this
region provides the least contaminated space for identifying
transition disk candidates, and we will correct our estimate for this
contamination.

Using the defined color regions we identify $\sim$5,600 candidate
HAeBe stars, $\sim$3,300 transition disk candidates, and $\sim$194,000
potential B- and A-type main-sequence stars.  Using the fraction of
classical Be stars found in the transition disk color space
($\sim$1/3) from \citet{Uzpen:2008}, we estimate that at most 2,200 of
these stars would be true transition disk objects.  The transition
disk fraction of 2,200/194,000 or 1\% is consistent with our prior
results. The fraction of pre-main-sequence objects, i.e., HAeBe and
transition disk objects, is ((5,600+2,200)/194,000)=4\%. This fraction
is higher than our value of 1.5$\pm$0.7\% within our complete working
sample.  However, our value is roughly consistent with the above
analysis given that the populations are likely to have some
contamination and given the differing levels of extinction used to
define the samples. Using these raw numbers we find that the
transition disk lifetime may be as long as one-third
(2,200/(5,600+2,200)) the primordial disk lifetime.  Using our
estimate of 4$\pm$2 Myr for the transition disk lifetime, this result
would imply that the primordial disk lifetime is 12$\pm$6
Myr. Although long, this value is consistent, within uncertainties,
with disk clearing studies.  Conversely, if we were to use our
estimated primordial disk lifetime of 5$\pm$2 Myr, we find that
transition disks survive 2$\pm$1 Myr.  Our result of 2$\pm$1 Myr is
longer than that estimated by \citet{Cieza:2007}. Our result is
consistent with other studies, which find transition disk objects
occur within 1--10 Myr (e.g., \citealt{Cieza:2007}), but is longer
than their estimated timescale for disk clearing of 0.4 Myr. In
summary, the possible omission of HAeBe stars from our complete
working sample is not likely to be a large effect, given that we
arrive at similar results for the transition disk lifetime and
primordial disk lifetime from an independent, albeit more uncertain,
approach.

\section{Comparison to Other Studies}

In order to place our results in context of circumstellar disk
evolution scenarios, we compare the derived fractions to those of other
infrared surveys for circumstellar disks using $IRAS$ and $Spitzer$.
This comparison allows us to determine if improved infrared
sensitivity and the larger volume surveyed by GLIMPSE places greater
constraints on mid-IR excess frequency. We also compare our results to
a compilation of all equivalent mass stars in $Spitzer$-studied young
star clusters ($<$ 10 Myr). Young star clusters contain both
primordial circumstellar disks and second generation dust disks. By
comparing equivalent-mass stars over a small age range, 1--10 Myr, we
can determine if the warm debris disk fraction and disk dissipation
times are consistent.

\subsection{$IRAS$}

\textit{IRAS} was flux-limited at 0.5 Jy in the 12 $\mu$m band. This
limit for an A5V star occurs at $V$=5.0. Applying Malmquist bias
corrections we find that the \textit{IRAS} survey was complete to a
distance of 44 pc for intermediate-mass stars.  There are 84 stars in
the Tycho-2 Spectral Catalog that fall within a complete flux-limited
$IRAS$ sample of B8--A5 spectral types with IV or V luminosity
class. We use this band to compare the warm debris disk fraction to
the $IRAS$ Catalog. One of the 84 \textit{IRAS} stars, $\beta$
Pictoris, has a 12 $\mu$m excess indicative of a mid-IR excess debris
disk or about 1.2$\pm$1.2$\%$.  Zeta Leporis, the other known $IRAS$
12 $\mu$m excess, does not appear in the Tycho-2 Spectral Catalog and
is omitted from this analysis.  The GLIMPSE survey is complete at
distances of 58--410 pc for intermediate-mass stars.  GLIMPSE surveyed
220 deg$^{2}$ of the celestial sphere or $220/41253$ of the solid
angle $IRAS$ covered. Therefore, the GLIMPSE survey is complete to an
equivalent sphere of 72 pc in radius.  This constitutes $\sim$4 times
the volume of the \textit{IRAS} survey. Our result for the frequency
of mid-IR debris disks, $0.3\pm0.3\%$, is consistent with the value
from the $IRAS$ survey, while encompassing four times the volume,
leading to smaller uncertainties.

Transition disks were not detected with $IRAS$ but are present in
$\sim$1$\%$ of our sample. The reason for this discrepancy may be due
to the relative volumes covered.  $IRAS$ was an all-sky survey, while
GLIMPSE focused more on the Galactic Plane. The GLIMPSE survey was
designed to observe the majority of high-mass Galactic star forming
regions and has identified a number of new Galactic clusters
(\citealt{Mercer:2005}; \citealt{Kobulnicky:2005};
\citealt{Strader:2008}).  $IRAS$ did not go deep enough to be complete
at the nearest star forming regions. Since the disk fraction is
correlated with age (e.g., \citealt{Rieke:2005}), one would
preferentially see more excesses in the younger Galactic Plane
population of the GLIMPSE survey. This could result in the increased
detection rate of transition disks in our working sample.

\subsection{Spitzer Studies of Young Star Clusters}

One of the major goals of $Spitzer$ was to increase our understanding
of stellar and circumstellar evolution in the context of planet
formation.  To achieve that goal, a number of General Observer,
Guaranteed Time Observer, and Legacy programs have observed stellar
clusters and star forming regions over a large range of ages. In order
to understand the process of circumstellar disk dissipation, at least
16 stellar clusters have been observed by $Spitzer$ over the 1--11 Myr
age range (summarized in \citealt{Hernandez:2008}). We culled from the
literature 165 B8--A5 stars for the following clusters and
associations: $\sigma$ Ori (8; \citealt{Hernandez:2007}), $\gamma$
Velorum (14; \citealt{Hernandez:2008}), NGC 2362 (2;
\citealt{Dahm:2007}), $\eta$ Cha (1; \citealt{Megeath:2005}), NGC 7160
and Tr 37 (18 and 10; \citealt{Sicilia-Aguilar:2006}), Orion OBIa and
OBIb (14 and 18; \citealt{Hernandez:2006}), IC 348 (7;
\citealt{Lada:2006}), Upper Sco (35; \citealt{Carpenter:2006}), NGC
1333 (3; \citealt{Gutermuth:2008}), and NGC 2244 (35;
\citealt{Balog:2007}). These clusters have $Spitzer$ IRAC and/or MIPS
data with spectral types in either the associated paper or other
related papers found using SIMBAD.  Twelve of these stars are
surrounded by primordial circumstellar disks.  Figure~\ref{decay}
shows the primordial disk fraction versus age (Myr) for these 12
clusters. This figure shows that the incidence of primordial disks is
zero at ages greater than 4 Myr, indicating that primordial disks
around intermediate-mass stars dissipate within the first 4--5 Myr.
By comparison, \citet{Hernandez:2008} Figure~11 shows that disk
dissipation around low-mass stars is a slower process, requiring
10--15 Myr.  This result is consistent with individual cluster studies
that find differing disk fractions for intermediate- and low-mass
members (e.g., \citealt{Carpenter:2006}).  We find disk dissipation
occurring within 4$\pm$2 Myr based on pre-main-sequence disk fractions
within a complete sample of intermediate-mass field stars is in
agreement with the disk dissipation rate found within cluster studies.

Using $Spitzer$ studies of young star clusters, \citet{Currie:2008a}
found that debris disk fractions increase at ages from 5--10 Myr and
peak in the 10--15 Myr range. We found 27 stars identified as debris
disk systems out of the 165 B8--A5 cluster stars in recent surveys,
yielding a debris disk fraction of 16$\pm$3$\%$, which is consistent
with other studies of debris disk fraction using data at $\lambda >
24$ $\mu$m (e.g., 33$\pm$19\% at ages $<$ 10 Myr;
\citealt{Rieke:2005}). However, only two of the 27 debris disk
candidates exhibit warm circumstellar excesses at [8.0], HD 36444
\citep{Hernandez:2006} and BD+31 641B \citep{Currie:2008b}. The
frequency of warm debris disks within young stellar clusters,
1.2$\pm$0.9$\%$, is consistent with our somewhat more restrictive
result of 0.3$\pm$0.3$\%$ from field stars. This result implies that
wide-area surveys containing millions of stars such as GLIMPSE,
GLIMPSE II, and GLIMPSE 3D may be ideal laboratories to identify more
of these objects, and with more of these unique objects, better
studies can be conducted to illuminate the astrophysics of this short
stellar evolutionary period.

\section{Conclusion}

We determined that the incidence of B8--A5 main-sequence stars with
warm debris disks is 0.3$\pm0.3$$\%$, and transition disks is
1.2$\pm$0.6$\%$ within a complete sample of the GLIMPSE, 2MASS, and
Tycho-2 Spectral Catalogs. The excess fraction measured here is much
lower than the 10--20\% of main-sequence stars that exhibit far-IR
excesses at $\lambda \geq 60$ $\mu$m (e.g.,
\citealt{Hillenbrand:2008}).  These mid-IR excess sources may be
members of yet-identified young stellar clusters. The rarity of
main-sequence stars that exhibit mid-, but not near-IR excesses
suggests that this stage of star formation is short-lived, consistent
with prior observational results (\citealt{Wolk:1996};
\citealt{Cieza:2007}).

We find the fraction of transition disk systems in our complete
working sample and a more inclusive biased sample are in agreement
with one another.  Investigation of likely transition and primordial
disks, using stellar models and statistical techniques, allows us to
constrain the lifetime of B8--A5 primordial circumstellar disks to
5$\pm$2 Myr using our complete working sample and 4$\pm$2 Myr using a
more inclusive sample that is not corrected for Malmquist bias.  These
values are consistent with those found by a growing number of studies
targeting young stellar clusters using $Spitzer$.  Our study provides
a complementary measurement using a significantly larger and
independently selected sample of intermediate-mass field stars.
Further investigation of these sources will help identify the place of
mid-IR excess circumstellar disks in the evolutionary process of star
and planet formation. However, our results do not provide strong
support for either current competing planet formation theories.

\acknowledgments

The authors would like to thank the anonymous referee for suggestions
that improved this paper. We kindly thank M.R. Meade, B.L. Babler,
R. Indebetouw, B. A. Whitney, C. Watson, and E. Churchwell for their
use of the GLIMPSE data reduction pipeline. We thank Dana Backman and
Lynne Hillenbrand for their helpful comments. B.U. acknowledges
support from a NASA Graduate Student Researchers Program fellowship,
grant NNX06AI28H.  This research has made use of the SIMBAD database,
operated at CDS, Strasbourg, France. This publication makes use of
data products from the Two Micron All Sky Survey, which is a joint
project of the University of Massachusetts and the Infrared Processing
and Analysis Center/California Institute of Technology, funded by the
National Aeronautics and Space Administration and the National Science
Foundation.

\clearpage

\begin{deluxetable}{lcccccccccc}
\tabletypesize{\footnotesize}
\rotate
\tablewidth{0pc}
\tablecaption{Stellar and Circumstellar Parameters of Disk Systems}
\tablehead{
\colhead{ID} &
\colhead{$K$-[8.0]} &
\colhead{[8.0]-[24]} &
\colhead{T$_{Disk}$} &
\colhead{$\frac{L_{IR}}{L_{*}}$} & 
\colhead{EW(H$\alpha$)$_{corr}$} &
\colhead{T$_{eff}$} &
\colhead{\textit{v} sin \textit{i}} &
\colhead{$V$} &
\colhead{Sp. Type} &
\colhead{Comment} \\
\colhead{} &
\colhead{[mag]} &
\colhead{[mag]} &
\colhead{(K)} &
\colhead{} &
\colhead{(\AA)} &
\colhead{(K)} &
\colhead{(km s$^{-1}$)} &
\colhead{[mag]} &
\colhead{} &
\colhead{} \\
\colhead{(1)} &
\colhead{(2)} &
\colhead{(3)} &
\colhead{(4)} &
\colhead{(5)} &
\colhead{(6)} &
\colhead{(7)} &
\colhead{(8)} &
\colhead{(9)} &
\colhead{(10)} &
\colhead{(11)} }
\startdata 
G305.4232--00.8229 & 0.79 & 1.27 & 675$^{+37}_{-33}$ & 0.0038$^{+0.0003}_{-0.0002}$ & -21.98$\pm$0.42 & 13520$\pm$510 & 230$\pm$20 & 9.23 & B6/8 V(E) & T \\ 
G307.9784--00.7148 & 0.40 & 1.06 & 557$^{+22}_{-16}$ & 0.00090$^{+0.00007}_{-0.00008}$ & -13.27$\pm$0.69 & 13220$\pm$520 & 240$\pm$20 & 7.92 & B8 V(N) & T \\
G311.0099+00.4156 & 0.37 & 2.79 & 315$^{+4}_{-3}$ & 0.0027$^{+0.0001}_{-0.0003}$ & 5.21$\pm$0.03 & 9800$\pm$130 & 235$\pm$25 & 8.07 & A3 IV & D \\
G311.6185+00.2469 & 0.28 & 1.82 &  306$^{+13}_{-6}$ & 0.0012$^{0.0001}_{-0.0001}$ & -10.90$\pm$0.23 & 10540$\pm$255 & 265$\pm$20 & 9.84 & B8/9 IV/V & T \\
G314.3136--00.6977 & 0.22 & 1.45 & 328$^{+11}_{-9}$ & 0.00096$^{+0.00007}_{-0.00008}$ & -12.63$\pm$0.29 & 9870$\pm$130 & 205$\pm$20 & 8.49 & A1 IV & T \\
G321.7868+00.4102 & 0.77 & 1.25 & 556$^{+18}_{-12}$ & 0.0021$^{+0.0002}_{-0.0001}$ & -26.81$\pm$0.53 & 12340$\pm$505 & 230$\pm$20 & 9.16 & B8VN & H \\
\enddata
\tablecomments{ (6) H$\alpha$ EW corrected for underlying stellar absorption. (11) T: Transition disk, D: Debris disk H: Herbig AeBe disk}
\label{des}
\end{deluxetable}

\clearpage

\begin{deluxetable}{cccc}
\tablewidth{0pc}
\tablecaption{Survey Completeness Limits}
\tablehead{
\colhead{Sp. Type} &
\colhead{$\sigma_{M}$\tablenotemark{1}} &
\colhead{m$_{comp}$\tablenotemark{2}} &
\colhead{D$_{Sp}$\tablenotemark{3}} \\
\colhead{} &
\colhead{[mag]} &
\colhead{[mag]} &
\colhead{[pc]} }
\startdata 
A5V & 0.60 & 10.33 & 440 \\
A4V & 0.79 & 10.05 & 410 \\
A3V & 0.67 & 10.15 & 460 \\
A2V & 0.79 & 9.99 & 460 \\
A1V & 0.65 & 10.08 & 470 \\
A0V & 0.65 & 10.05 & 480 \\
B9V & 0.69 & 9.83 & 500 \\
B8V & 0.69 & 9.95 & 640 \\
\enddata
\tablenotetext{1}{Measured dispersion in absolute magnitude from
  $Hipparcos$ parallax}
\tablenotetext{2}{Measured m$_{V}$ at which our
  sample is complete with assumed reddening}
\tablenotetext{3}{The distance at which our sample is complete.}
\label{comp}
\end{deluxetable}

\clearpage

\begin{deluxetable}{ccccc}
\tablewidth{0pc}
\tablecaption{The Complete Working Sample}
\tablehead{
\colhead{Sp. Type} &
\colhead{N$_{V}$\tablenotemark{1}} &
\colhead{N$_{IV}$\tablenotemark{2}} &
\colhead{N$_{V}$$_{IRAS}$\tablenotemark{3}} &
\colhead{N$_{IV}$$_{IRAS}$\tablenotemark{4}} }
\startdata 
A5 & 4 & 8 & 5 & 5 \\
A4 & 1 & 3 & 1 & 1 \\
A3 & 11 & 9 & 9 & 4 \\
A2 & 14 & 12 & 5 & 5 \\
A1 & 33 & 23 & 12 & 0 \\
A0 & 63 & 19 & 18 & 7 \\
B9 & 55 & 38 & 6 & 0 \\
B8 & 25 & 20 & 6 & 0 \\
\enddata
\tablenotetext{1}{Measured number of luminosity class V
  GLIMPSE sources within the complete sample}
\tablenotetext{2}{Same as (1) for luminosity
 class IV or IV/V GLIMPSE sources}
\tablenotetext{3}{Measured number
of luminosity class V sources within the complete $IRAS$ sample} 
\tablenotetext{4}{Same as
(3) for luminosity class IV or IV/V $IRAS$ sources}
\label{types}
\end{deluxetable}

\clearpage

\begin{deluxetable}{ccccc}
\tablewidth{0pc}
\tablecaption{Sample Statistics}
\tablehead{
\colhead{Sample} &
\colhead{DF$_{WD}$} &
\colhead{DF$_{T}$} &
\colhead{DF$_{P}$} &
\colhead{Lifetime$_{P}$}  \\
\colhead{} &
\colhead{$\%$} &
\colhead{$\%$} &
\colhead{$\%$} &
\colhead{Myr} }
\startdata 
Complete Working & 0.3$\pm$0.3 & 1.2$\pm$0.6 & 1.5$\pm$0.7 & 5$\pm$2  \\
$V$$\leq$9.25 & 0.6$\pm$0.6 & 1.8$\pm$1.1 & 2.5$\pm$1.2 & 9$\pm$4  \\
Biased & 0.2$\pm$0.2 & 0.8$\pm$0.4 & 1.0$\pm$0.5 & 4$\pm$2  \\
\enddata
\tablecomments{See text for sample characteristics and
descriptions. DF (Disk Fraction); WD (warm debris disk); T (transition disk); P (Primordial disk)}
\label{sum}
\end{deluxetable}

\clearpage

\begin{deluxetable}{ccc}
\tablewidth{0pc}
\tablecaption{Longevity of Spectral Type}
\tablehead{
\colhead{Sp. Type} &
\colhead{N$_{Stars}$\tablenotemark{1}} &
\colhead{Age\tablenotemark{2}} \\
\colhead{} &
\colhead{} &
\colhead{Myr} }
\startdata 
A5 & 12 & 980 \\
A4 & 4 & 750 \\
A3 & 20 & 750 \\
A2 & 26 & 530 \\
A1 & 56 & 430 \\
A0 & 82 & 330 \\
B9 & 93 & 220 \\
B8 & 45 & 150 \\
\enddata
\tablenotetext{1}{Measured number of luminosity
  class V and IV GLIMPSE sources}
\tablenotetext{2}{Stellar lifetime of
  \citet{Siess:2000} models in Myr for nearest estimated mass to
  measured spectral type}
\label{tageest}
\end{deluxetable}

\clearpage

\begin{deluxetable}{ccc}
\tablewidth{0pc}
\tablecaption{The $V$$\leq$9.25 Sample}
\tablehead{
\colhead{Sp. Type} &
\colhead{N$_{V}$\tablenotemark{1}} &
\colhead{N$_{IV}$\tablenotemark{2}}  }
\startdata 
A5 & 2 & 3 \\
A4 & 1 & 0 \\
A3 & 5 & 3 \\
A2 & 8 & 8 \\
A1 & 15 & 4 \\
A0 & 31 & 10 \\
B9 & 35 & 19 \\
B8 & 14 & 5 \\
\enddata
\tablenotetext{1}{Measured number of luminosity class V GLIMPSE
  sources within the $V$$\leq$9.25 sample}
\tablenotetext{2}{Measured number of
  luminosity class IV or IV/V GLIMPSE sources within the $V$$\leq$9.25
  sample}
\label{types2}
\end{deluxetable}

\clearpage

\begin{figure}
\hbox{
    \psfig{file=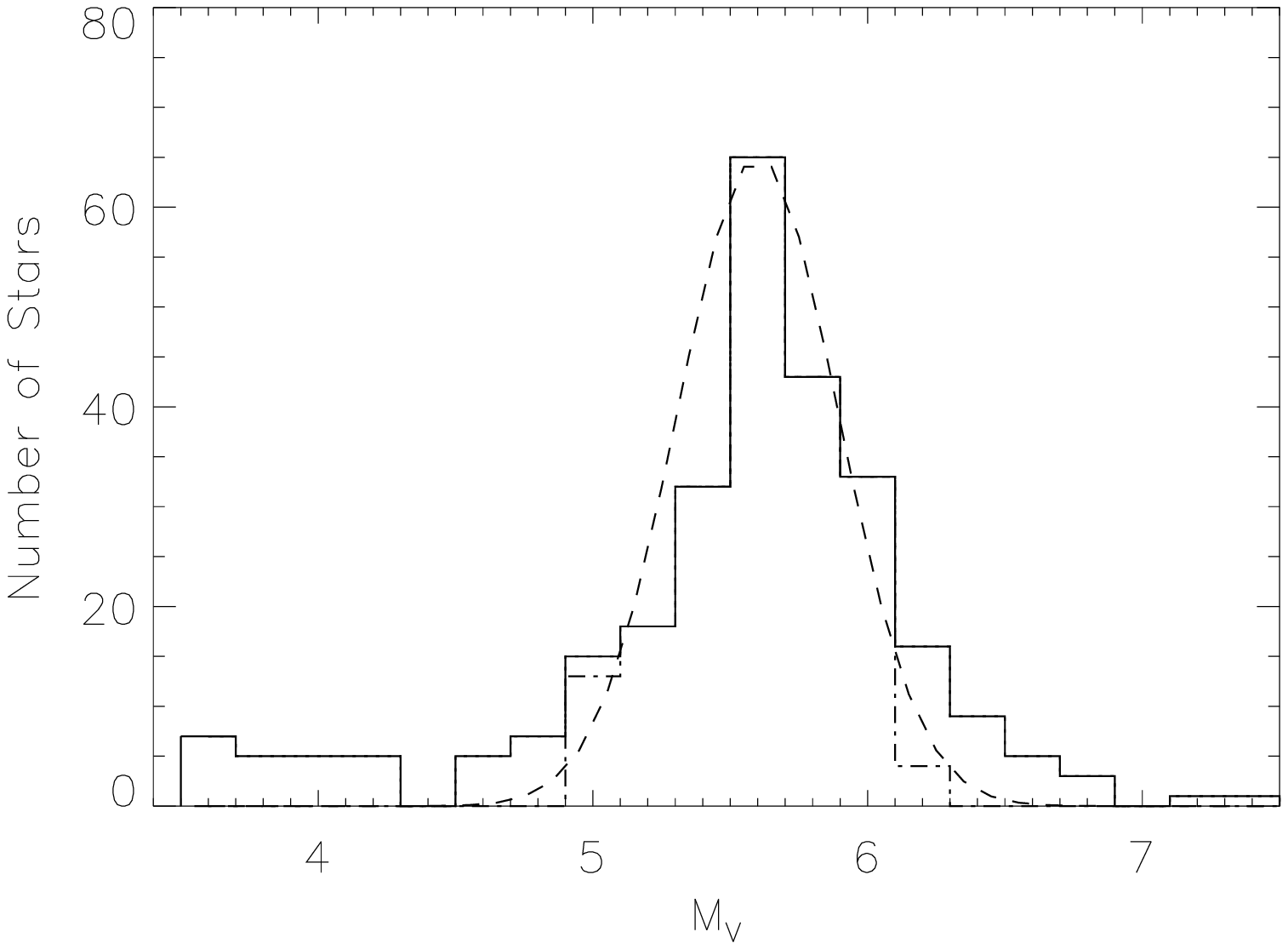,width=4.0in}
}
\hbox{
    \psfig{file=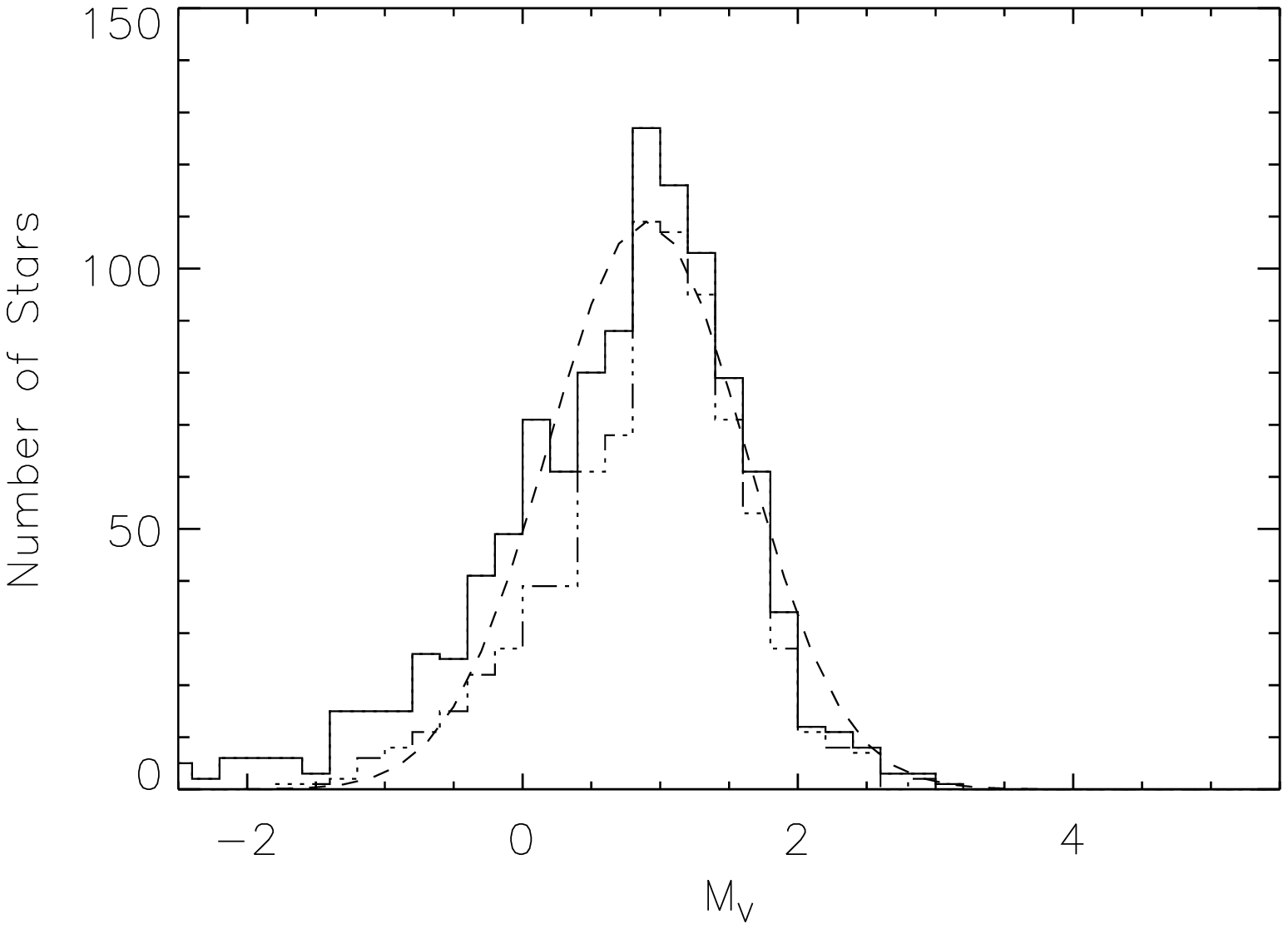,width=4.0in}
}\caption{($top$) The distribution of absolute magnitude for 208 K0V
stars with $Hipparcos$ parallaxes at an S/N $>$ 3. The dashed curve is
a Gaussian distribution with a dispersion of 0.29, centered at 5.62,
and a maximum equal to the maximum of the K0V histogram. The
distribution is nearly Gaussian.  ($bottom$) The distribution of absolute
magnitude for 772 A0V stars with $Hipparcos$ parallaxes at an S/N $>$ 3
in the All-Sky Catalog. The dashed curve is a Gaussian distribution
with a dispersion of 0.65, centered at 0.92, and a maximum equal to
the maximum of the A0V histogram. The A0V histogram is slightly
asymmetric, resulting in a greater dispersion. Similar analyses were
performed for all spectral types B8--A5 in order to determine the
dispersion in absolute magnitude among each spectral type.}
\label{dist}
 \end{figure}

\clearpage

\begin{figure}
    \psfig{file=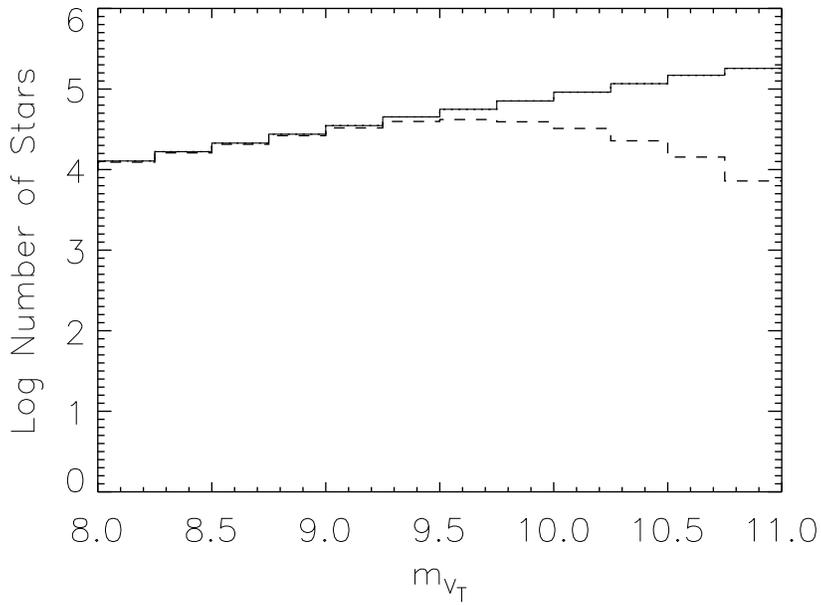,width=5.0in}
    \caption{Histograms of Tycho visual magnitude for both the Tycho-2
    Catalog ($solid$ $line$) and Tycho-2 Spectral Catalog ($dashed$ $line$).
    The Tycho-2 Spectral Catalog derives its photometric data from the
    Tycho-2 Catalog and starts to drop off in completeness after 9.25
    magnitudes. Therefore, all stars fainter than m$_{V}$=9.25 may not
    have spectral information determined yet.}
    \label{sphist}
\end{figure}

\clearpage 

 \begin{figure}
\hbox{
    \includegraphics[width=4.0in]{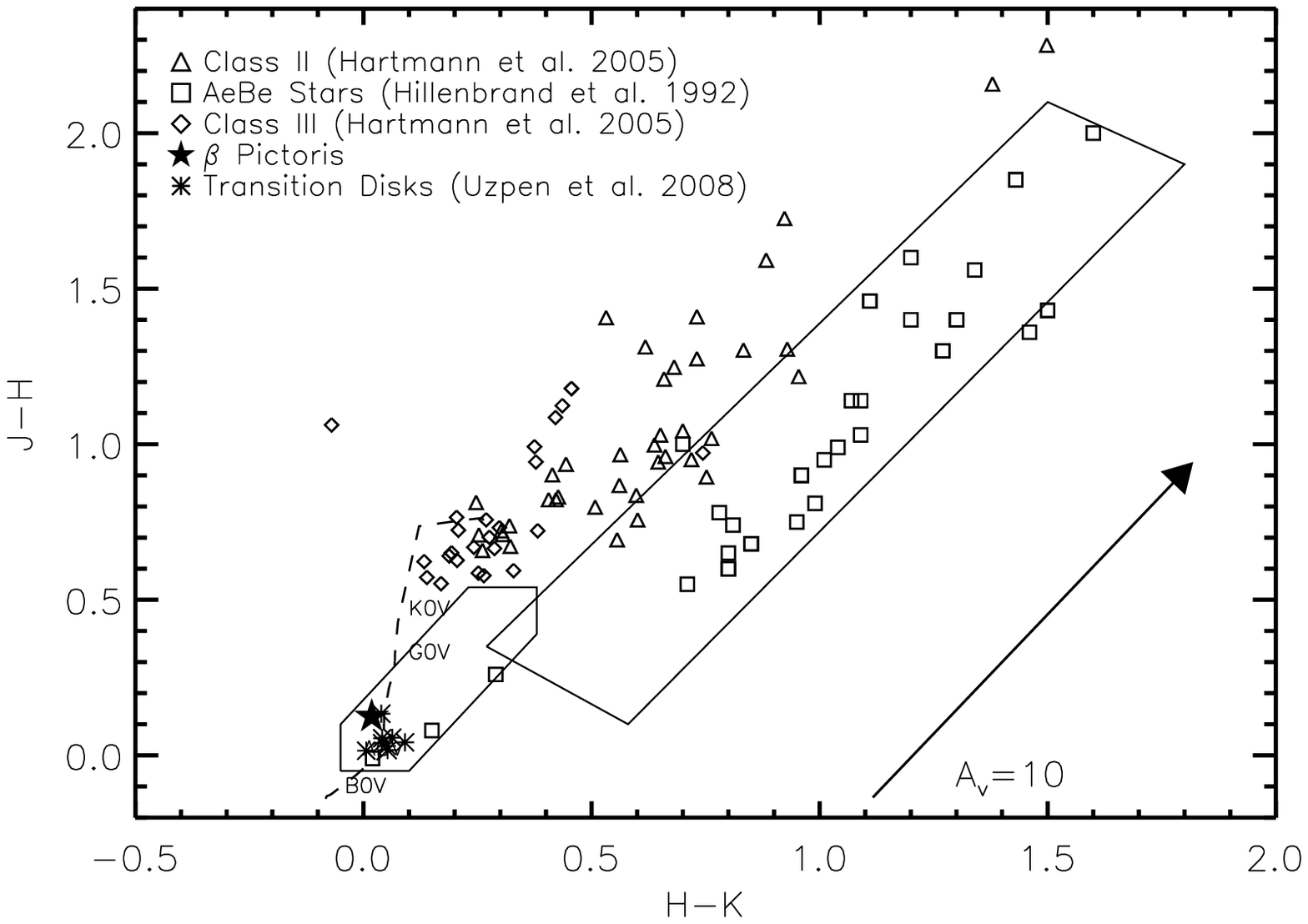}
}
\hbox{
    \includegraphics[width=4.0in]{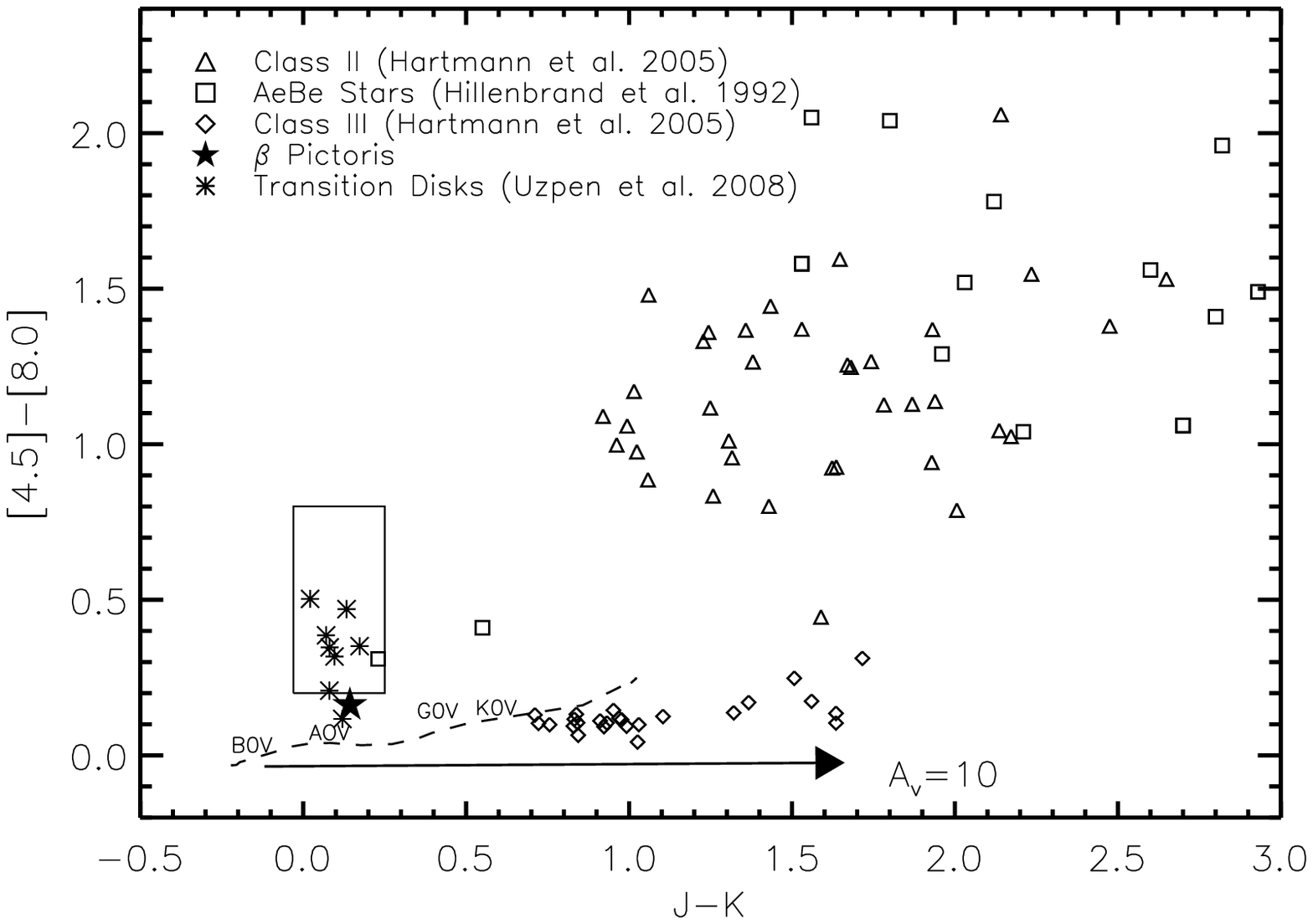}
}
 \caption{($top$) $J$--$H$ vs. $H$--$K$ color diagram. Low-mass
pre-main-sequence stars are shown by triangles and diamonds.  HAeBe
stars are shown by squares and lie in a region that reddened normal
main-sequence stars cannot occupy.  The fiducial main-sequence is
shown by the dashed line.  The parallelogram is used to define the
region in which we expect to find HAeBe stars. The lower hexagonal
region encompasses unreddened B and A main-sequence stars to those
with A$_{V}$=4. ($bottom$) [4.5]--[8.0] vs. $J$--$K$ color
space. Gaseous pre-main-sequence stars clearly lie above and to the
right of the main-sequence. Evolved low-mass pre-main-sequence stars
lie to the right of the main-sequence, exhibiting colors consistent
with reddening only. Intermediate-mass stars exhibiting excesses
consistent with transition disks ($asterisks$) occupy a unique region
of color space outlined by the rectangular box. Stars within this box
are not highly reddened, exhibit a mid-IR excess, and may be
undergoing disk clearing.}
    \label{2color}
\end{figure}

\clearpage

\begin{figure}
    \psfig{file=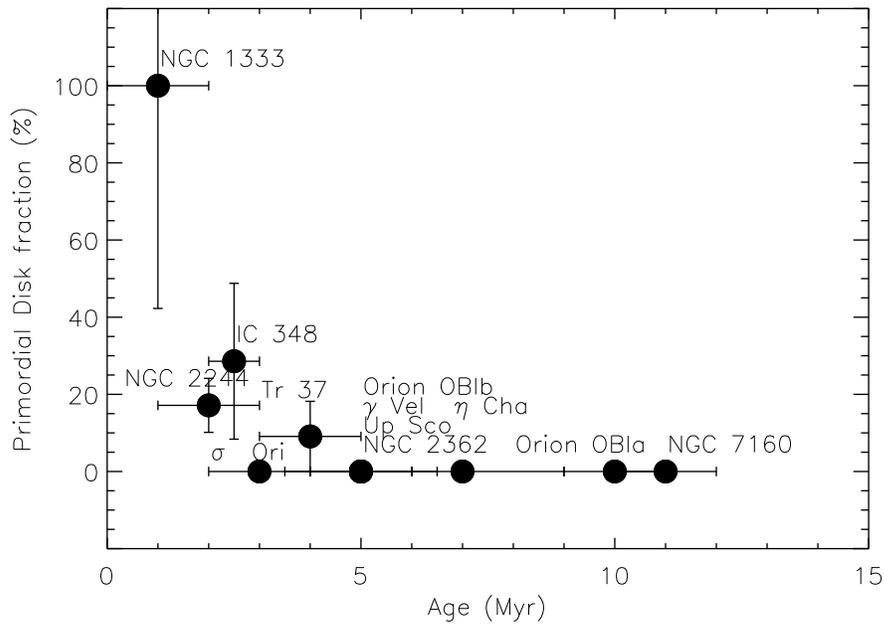,width=5.0in}
    \caption{Cluster age versus primordial disk fraction for $Spitzer$-observed 
    clusters. None of the clusters older than 4 Myr contain
    any primordial disks in the spectral range B8--A5. The decay rate
    is significantly faster than that for cluster members of all
    masses shown in \citet{Hernandez:2008} Figure~11. Orion OBIb,
    $\gamma$ Vel, Up Sco, and NGC 2362 are 5 Myr old and contain no
    primordial disks, therefore their points overlap.}
    \label{decay}
\end{figure}

\clearpage

\bibliography{apj-jour,paper4} 
\end{document}